\documentclass[12pt]{article}

\usepackage{amssymb,latexsym,epsfig}
\usepackage{cite}

\newcommand{\sect}[1]{\setcounter{equation}{0}\section{#1}}
\newcommand{\subsect}[1]{\subsection{#1}}

\newfont{\extra}{msbm10 scaled\magstep1}

\def\be{\begin{equation}}
\def\ee{\end{equation}}
\def\bea{\begin{eqnarray}}
\def\eea{\end{eqnarray}}

\def\bx{\beta_x}
\def\bt{\beta_t}

\def\la{\lambda}

\def\q{$q$--}
\def\qx{q_x}
\def\px{\partial_x}
\def\pxx{\partial_{xx}^2}
\def\pt{\partial_t}
\def\s{{\rm s}}

\def\N{\mathbb N}

\def\Z{\mathbb Z}

\begin{document}

\begin{center}
{\LARGE{\bf{Discrete $q$--derivatives \\  and \\[0.45cm] symmetries 
of $q$--difference
equations}}}
\end{center}

\bigskip\bigskip

\begin{center}
D. Levi $^1$, J. Negro $^2$ and M.A. del Olmo $^2$
\end{center}

\begin{center}
$^1${\sl Dipartimento di Fisica, Universit\'a  Roma Tre and
INFN--Sezione di
Roma Tre \\
Via della Vasca Navale 84, 00146 Roma, Italy}\\
\medskip

$^2${\sl Departamento de F\'{\i}sica Te\'orica, Universidad de Valladolid, 
\\
E-47011, Valladolid, Spain.}\\
\medskip

{e-mail:levi@fis.uniroma3.it,  jnegro@fta.uva.es, olmo@fta.uva.es}
\end{center}

\vskip 1.5cm
\centerline{\today}
\vskip 1.5cm

\bigskip

\begin{abstract}
In this paper we extend the {\it umbral calculus}, developed to deal  
with difference equations on uniform lattices, to
$q$-difference equations. We show that many of the properties
considered for shift invariant difference
operators satisfying the umbral calculus can be implemented to the
case of the $q$-difference operators. This
{\it $q$-umbral calculus} can be used to provide solutions to linear
$q$-difference equations and $q$-differential delay equations. To illustrate
the method, we will apply the obtained results to the construction of
symmetry solutions for the $q$-heat equation and to solve a linear ordinary 
second order $q$-difference equation.

\end{abstract}
\vskip 1cm

\vfill
\eject

\sect{Introduction\label{introduccion}}

$Q$-functions appear in many physical problems. They enter  
in the study of exactly solvable models in statistical mechanics \cite{baxter}, 
in  conformal field theory \cite{Fadeev}, and are thus very relevant 
for applications. For example, $q$-exponential distributions can be obtained 
following Gibbs' procedure from the stationary conditions on a certain generalized 
entropy \cite{NEXT}. Standard \q exponential functions are also used to extrapolate 
between Fermi--Dirac ($q = \infty$) and Bose--Einstein ($ q=0$) statistics, 
passing through Maxwell--Boltzmann ($q=1$) statistics \cite{anally}.

In the case of difference equations one had proved \cite{lno01} that there 
exists a very powerful method
for systematically discretizing linear differential equations 
while preserving their properties.
Here we extend  that method  to
the case of $q$-difference equations. We can show  that many of
the properties considered in \cite{lno01} for shift invariant
difference operators satisfying the umbral calculus 
\cite{umbral1,umbral2,umbral3,umbral4,umbral5,aizawa} can be extended to
the case of the $q$-difference operators considered in
Ref.\cite{floviq,floviq2}. For any $q$-difference operator this
$q$-umbral calculus can be 
applied to provide solutions to linear $q$-difference equations
and $q$-differential delay equations. As an illustration, we will
apply the method in the construction of symmetry solutions for the $q$-heat
equation and to a linear ordinary $q$-difference equation.

The paper is organized as follows. In Section 2 we define a $q$-difference
equation in
$p$ independent variables and, for the sake of simplicity,  just one
dependent variable and characterize the symmetry transformations
which will leave the equation invariant. Section 3 is devoted to
the study of the properties of $q$-calculus, showing the
differences and the similarities between differential and
$q$-difference calculus. In particular in Section 4 we will
discuss from an analytic and numerical point of view the simplest
$q$-functions which will be used later, in Section 5, along with a few
examples. In Section 5 we present the symmetries of a $q$-heat equation
and solve a differential
$q$-delay equation using the correspondence between $q$-calculus and
differential calculus. Section 6 is dedicated to  few conclusive remarks.

\sect{$q$-difference equations and its Lie symmetries}\label{liesimetrias}

Let us consider a linear $q$-difference equation, involving,
for notational simplicity, only one scalar function  $u(x)$ of $p$
independent variables $x=(x_1,x_2,\dots,x_p)$ evaluated at a
finite number of points on a lattice. Symbolically we write
\bea\label{ecuaciondiscreta} E_N(x,\ T^{a}u(x),\
T^{a_{i_1}}\Delta_{x_{i_1}} u(x),\
T^{a_{i_1i_2}}\Delta_{x_{i_1}}\Delta_{x_{i_2}} u(x), \dots, \\
\nonumber
 T^{a_{i_1i_2 \dots i_N}}\Delta_{x_{i_1}}\Delta_{x_{i_2}}\dots 
\Delta_{x_{i_N}} u(x) )=0 , \qquad
a=(a_1,a_2,\dots, a_p),
\eea
where $ a, a_{i-1}, a_{i_1 i_2}, \dots $ are multi-indices, 
$E_N$ is some given function of its arguments, $i_1, i_2, \dots, 
i_N$ take values between $1$ and $p$. We use the shortening notation 
$T^{a}u(x) =\{ T^{a_1}_{x_1}T^{a_2}_{x_2}\cdots T^{a_p}_{x_p}\,
u(x)  \} ,$
where $a_i$, with $i=1,2,\dots, p$, takes values between $m_i$ and $n_i$,
with $\ m_i, \ n_i$  fixed integers
($m_i \leq n_i$), and  the individual \q shift operator $T^{a_i}_{x_i}$ is
given by 
\be \label{1} T^{a_i}_{x_i}u(x)=
u(x_1,x_2,\dots,x_{i-1}, q_i^{a_i} x_i, x_{i+1},\dots,x_p). 
\ee
The other $q$-shift operators $T^{a_{{i_1i_2}}}, \dots$  are
defined in a similar way. The operator $\Delta_{x_i}$ is a
$q_i$-difference operator which in the continuous limit when $q_i
\rightarrow 1$ goes into the partial derivative with respect to
the $x_i$ variable (few examples of it are given in Section
3). By $q_i$ we denote  a positive parameter which defines a  non
uniform lattice of the variable $x_i \ (i=1,\dots, p)$.

To study the symmetries 
of eq.(\ref{ecuaciondiscreta})  we will  use the
approach introduced in Ref. \cite{decio}, based on the formalism
of evolutionary vector fields for differential equations
\cite{olver}. As  in the case of 
differential equations the symmetry group of a discrete equation
is characterized  by those transformations of the equation that carry
solutions $u(x)$ into
solutions $\tilde u(x)$.  Moreover we look only for those
symmetries which in the continuous limit go over to Lie point
symmetries. In such a case  the 
infinitesimal symmetry  generators of  the symmetry group of equation
(\ref{ecuaciondiscreta})  in evolutionary form have the general
expression
\bea \label{vectorgeneral}
X_e\equiv Q(x,u)\partial_u &=& 
\left( \sum_{i=1}^{p} \xi _i (x, T^{a}u, \{ q_j\}_{j=1}^p ) T^{b}
\Delta_{x_i} u -\phi(x,T^{c}u,\{ q_j\}_{j=1}^p )
\right)\partial_u ,
\eea
with  $\xi _i (x, T^{a}u,\{ q_j\}_{j=1}^p )$ and $\phi(x,T^{c}u,\{ q_j\}_{j=1}^p )$ such that
 in the continuous limit go over to $\xi _i (x, u)$ and $\phi(x, u)$, 
the infinitesimal generators of the 
 corresponding Lie point
symmetries.  The group transformations are obtained by
integrating the differential equation \be \label{2} \frac {d
\tilde u(\tilde x)}{d g} = Q(\tilde x, \tilde u(\tilde x)),
\qquad \tilde u(\tilde x, g = 0) = u(x), \ee where $g$ is the
group parameter. Eq.(\ref{2}) can be integrated on the
characteristics in the continuous limit, however in the discrete
case the integration of the corresponding differential \q difference 
equation can almost never be carried out.

Eq. (\ref{ecuaciondiscreta}) is of order $N$ in the
difference operators and of order $i_1 + i_2 + \dots + i_N$
in the shift operators. Hence, constructing the following 
prolongation of $X_e$ 
\be\label{prolongationvector} pr
X_e=\sum_a T^a Q\partial_{T^a u} + \sum_{i_1}
T^{a_{i_1}}Q^{x_{i_1}}\partial_{T^{a_{i_1}}\Delta_{x_{i_1}}u(x)}
+ \sum_{i_1,i_2}
T^{a_{i_1i_2}}Q^{x_{i_1}x_{i_2}}\partial_{T^{a_{i_1i_2}}
\Delta_{x_{i_1}}\Delta_{x_{i_2}}u(x)}  + \dots \ee
we find that the invariance condition
\be\label{prolongacion} pr X_e E_N |_{E_N=0}=0 
\ee 
must be satisfied.
The summations
in (\ref{prolongationvector}) are over all the sites present in
(\ref{ecuaciondiscreta}). By $Q^{x_{i_1}},\quad
Q^{{x_{i_1}}{x_{i_2}}},\dots$ we denote the  total variations of
$Q$, i.e.,
$$
 Q^{x_{i_1}}=\Delta_{x_{i_1}}^T Q, \qquad
 Q^{{x_{i_1}}{x_{i_2}}}=\Delta_{x_{i_1}}^T \Delta_{x_{i_2}}^T Q ,\qquad 
\cdots,
$$
where, in the simple case of the $q$-right derivative $\Delta_x^+$ 
presented  below in eq.(\ref{qderivadas}), the
partial variation
$\Delta_{x_{i_1}}$ is defined by
$$\begin{array}{ll}
\Delta_{x_{i_1}} f(x_1, \dots , x_p, u(x_1, \dots , x_p), 
\Delta_{x_{j}}u(x_1, \dots , x_p), \dots)
\\[0.2cm] \nonumber
\quad = \frac 1{(q_{i_1}-1) x_{i_1}} [f(x_1,
\dots q_{i_1} x_{i_1}, \dots, x_p, u(x_1, \dots , x_p), 
\Delta_{x_{j}} u(x_1, \dots , x_p), \dots)  \\
[0.2cm] \nonumber \qquad   -f(x_1, \dots , x_p, u(x_1, \dots , x_p), 
\Delta_{x_{j}}u(x_1, \dots , x_p),
\dots)] ,
\end{array}$$
and the total variation $\Delta_{x_{i_1}}^T$ by
$$\begin{array}{ll}
\Delta_{x_{i_1}}^T f(x_1, \dots , x_p, u(x_1, \dots , x_p),
\Delta_{x_{j}}u(x_1, \dots , x_p), \dots)= \\ [0.2cm] \nonumber
\frac 1{(q_{i_1}-1) x_{i_1}} [f(x_1, \dots q_{i_1} x_{i_1},
\dots, x_p, u(x_1, \dots q_{i_1} x_{i_1}, \dots, x_p),
\Delta_{x_{j}} u(x_1, \dots q_{i_1} x_{i_1}, \dots, x_p), \dots)
\\ [0.2cm] \nonumber  \quad -f(x_1, \dots , x_p, u(x_1, \dots , x_p), 
\Delta_{x_{j}} u(x_1, \dots , x_p),
\dots)] .
\end{array}$$
Notice that expressions
(\ref{vectorgeneral}--\ref{prolongationvector}) are analogous
to those of the continuous case \cite{olver} and can be derived
in a similar way \cite{decio}.

The group generated by the prolongations also transforms solutions into
solutions, and  $\Delta_{x_i}u,\Delta_{x_i}\Delta_{x_j}u, \dots $ 
(up to order $N$) into the variations of  $\tilde u$ with respect to the
corresponding $\tilde x_i$.

The symmetries of  equation (\ref{ecuaciondiscreta}) are given by
condition (\ref{prolongacion}), which give rise  to a set of
determining equations for $\xi _i$ and $\phi$  obtained as
coefficients of the linearly independent expressions in the
discrete derivatives $\Delta_{x_i}u$,
$\Delta_{x_ix_j}u$,).

The Lie commutators of the vector fields $X_e$ are obtained by
commuting their first prolongations and projecting them onto the
symmetry algebra $\cal G$, i.e.,
\bea\label{comnmutador} [X_{e_1},X_{e_2}] &=&
[pr^{1}X_{e_1},pr^{1}X_{e_2}]|_{\cal G}\\ [0.2cm]\nonumber 
&=&\left( Q_1\frac{\partial Q_2}{\partial u} -Q_2\frac{\partial
Q_1}{\partial u} + Q^{x_i}_1\frac{\partial Q_2}{\partial u_{x_i}}
-Q_2^{x_i}\frac{\partial Q_1}{\partial u_{x_i}}  \right
)\partial_u \ , \eea where the $\partial u_{x_i}$ terms disappear
after projection onto $\cal G$.

The formalism presented above  may become quite involved, but the
situation is simpler for linear equations where we can use a
reduced Ansatz. In this  case   we can assume that the
evolutionary vectors (\ref{vectorgeneral}) have the form
\be\label{vectorrestringido} X_e= \left( \sum_i \xi _i (x, T^{a},
q_j) \Delta_{x_i} u -\phi(x,T^{a}, q_j) u \right)\partial_u . \ee
The vector fields $X_e$ can be written as $X_e= (\hat X u)
\partial_u$ with \be\label{vectorlie} \hat X=  \sum_i \xi _i (x,
T^{a}, q_j) \Delta_{x_i}  -\phi(x,T^{a}, q_j) .
\ee 
Notice that $\hat X$ span a subalgebra of the whole
Lie symmetry algebra (see Ref. \cite{decio}).

 If the system is  nonlinear the simplification (\ref{vectorrestringido}) is
too restrictive and is almost impossible to get a non trivial result
since the number of terms to consider is a priori infinite.

\sect{ \q Calculus}\label{umbral}

{\ In this section we  present the generalities of \q
calculus \cite{kac}.} We will restrict ourself for the sake of
simplicity to one independent variable. Moreover, in the
following we will consider just the
simplest \q derivatives (at the right, at the left and
symmetric, respectively) \bea\label{qderivadas}
&& \Delta_{x}^{+} = \frac 1{q_x^+ x} (T_x -1),\\[0.3cm]
&& \Delta_{x}^{-} = \frac 1{q_x^- x} (1-T_x^{-1}),\\[0.3cm]
&& \Delta_{x}^{{\s}} = \frac 1{q_x^\s x}(T_x -T_x^{-1}) , \label{3.3} \eea
where $q_x$ is a real dilation positive parameter associated to the variable
$x$,  $q^i_x, i=\pm,s$ are  given by \be q_x^+=q_x -1 \qquad
q_x^-=1-\frac{1}{q_x} \qquad q_x^\s=q_x -\frac{1}{q_x},
\ee 
and $T_x$ is a \q dilation
\be\label{translation} T_x f(x)= f(q_x x)\qquad T_x^{-1} f(x)=
f({x}/{q_x}). 
\ee 
The  operator $T_x$ is the one-dimensional reduction of  eq. (\ref{1}). Formally we have 
\be\label{translationform} 
T_x= e^{(q_x - 1) x\partial_x} .
\ee 
When we do not
specify which \q derivative we are
using we will write just $\Delta_{x}$.

It is easy to see that, due to the form of the \q derivative
considered (\ref{qderivadas} - \ref{3.3}), the shift operator and the \q
derivatives do not commute. So the \q derivative operators are 
not shift invariant operators and do not satisfy
one of the basic conditions in the umbral calculus
\cite{umbral1,umbral2,umbral3,umbral4}.
We can, however, carry out an umbral calculus even if our \q
derivative are not shift invariant that we will call  {\em \q umbral calculus}. 
The \q umbral calculus is defined in the same way as the standard umbral
calculus, but  we do not require the commutativity of the delta
operators with the shift operators. The absence of this property has no consequence in
all the results presented in the following.

We can easily find that
\be\label{commutador}
[\Delta_{x},x]=\left\{ \begin{array}{lll} T_x \quad &{\rm for}\quad
&\Delta_{x}^{+}\\[0.25cm]
T_x^{-1} \quad&{\rm for}\quad  &\Delta_{x}^{-} .\\[0.25cm]
\frac{1}{1+q_x}(q_x T_x +T_x^{-1} )\quad &{\rm for}\quad  &\Delta_{x}^{\s}
\end{array}\right.\ee

If instead of the standard commutator, we consider the \q commutator 
defined as $[A,B]_{q^+}=A B -q A B$ and $[A,B]_{q^-}=A B -(1/q) A B$,
we have
\be\label{qcommutador}
[\Delta_{x},x]_{q}=1.
\ee
The result (\ref{qcommutador}) is not valid in the case of
the symmetric \q derivative. 
Moreover, the expression (\ref{qcommutador}) does not satisfy 
the Leibniz rule and, thus, the expression $[\Delta_{x}, x^{n}]_{q}$ 
becomes more and more complicate when we consider
$n= 2, 3, \cdots$ . So, here in the following we will consider just the 
standard
commutator.

We can define an operator $\bx$, depending on $T_x$, in the
spirit of umbral calculus \cite{umbral4,umbral5}, such that 
\be 
[\bx ,
T_x]=0,\qquad
 [\Delta_{x}, \bx x]=1 .
\ee
For the three \q derivatives introduced above
(\ref{qderivadas}) we can find the following explicit expressions of $\bx$:
\bea \nonumber
&& \bx ^+ = (\qx-1) x \px (T_x-1)^{-1}= q_x^+ x \px (T_x-1)^{-1} ,\\[0.25cm] 
\label{3.7}
&& \bx ^- = (1-\frac{1}{q_x}) x \px (1-T_x^{-1})^{-1}= q_x^- x \px
(1-T_x^{-1})^{-1},\\[0.25cm] \nonumber
&& \bx ^\s = (q_x -\frac{1}{q_x}) x \px (T_x-T_x^{-1})^{-1}=
q_x^\s x \px (T_x-T_x^{-1})^{-1} .\eea
These may not be the
only possible definitions.

It is easy to prove that always 
\be\label{tres.doce.bis}
\bx x\Delta_{x}= x \px.
\ee
Moreover, due to the presence of the $\px$ operator in
the definition of $\bx$, the \q umbral correspondence of an
explicit $x$--dependent differential equation will give rise to
a \q differential delay equation \cite{delay}.

We can reexpress the functions $\bx$ as an infinite series
in terms of the shift operators, thus proving that they commute
with the shift operators. From (\ref{translationform}) we get
 \be
x\px=\frac1 {\qx - 1} \ln T_x = \frac1 { \qx - 1} \ln (1 + (T_x
-1)) = \frac1 { \qx - 1} \sum_{n=1}^{\infty} (-1)^{n+1}\frac{(T_x
-1)^n}{n} .
\ee
Consequently  
\be 
\bx ^+=\frac{\qx ^+} { \qx - 1} \sum_{n=0}^{\infty}
(-1)^{n}\frac{(T_x -1)^n} {n+1} . \ee Similarly from
$T_x^{-1}=\exp{(-(\qx  - 1) x\px})$
\be
x\px=-\frac1 { \qx - 1 } \ln T_x^{-1} = -\frac1 {\qx - 1} \ln (1 +
(T_x^{-1} -1)) = \frac1{\qx - 1} \sum_{n=1}^{\infty}
(-1)^{n}\frac{(T_x^{-1} -1)^n}{n} \ee
we get 
\be \bx ^-=\frac{\qx
^-} {\qx - 1} \sum_{n=0}^{\infty} (-1)^{n}\frac{(T_x^{-1} -1)^n}
{n+1} =\frac{\qx ^-} { \qx - 1} \sum_{n=0}^{\infty}
\frac{(1-T_x^{-1} )^n} {n+1} . \ee
Finally, for the symmetric
derivative, as
\be \frac{T_x-T_x^{-1}}2=\sinh ((\qx - 1) x\px )
\ee we find that 
\be x\px=\frac1 {\qx - 1} \sinh ^{-1}\frac{T_x
-T_x ^{-1}}{2} = \frac1{\qx - 1} \sum_{n=1}^{\infty}
(-1)^{n+1}C_n\frac{(T_x - T_x^{-1})^{2n-1}}{2^{2n-1}} , 
\ee 
where
the coefficients $C_n$ are  given by
\be C_1=1 ,\qquad
C_n=\frac{\prod _{k=2}^n{(2k - 3)}}{(2 n - 1)\prod _{k=2}^n{(2 k
- 2)}} , \qquad \forall n \geq 2 . 
\ee 
So, we have: 
\be \bx
^\s=\frac{\qx ^\s} {\ln \qx} \sum_{n=0}^{\infty}
(-1)^{n}C_{n+1}\frac{(T_x - T_x^{-1})^{2n}} {2^{2n+1}} . 
\ee

For all functions $f$ and $g$, entire in $\beta_x x$, 
the Leibniz rule takes the form \be\label{leibnizrule}
[\Delta_{x}, f g]=[\Delta_{x},f] g + f [\Delta_{x},g] . \ee Note
that the  expression of Leibniz's rule (\ref{leibnizrule}) is exactly 
the same as for discrete derivatives \cite{lno01}. For the sake of
brevity we will write $D_x f := [\Delta_{x},f]$. The 
expression (\ref{leibnizrule}) is operatorial. If we want to have
a functional expression we need to project it  by acting on a
constant function as $1$. In this case, however, the Leibniz rule 
(\ref{leibnizrule}) is an identity.

Taking into account the Leibniz rule (\ref{leibnizrule}) 
we can prove for the three \q
derivatives introduced above the following property \be [\Delta_{x},(\bx x)^n] =
D_x (\bx x)^n = n (\bx x)^{n-1},\qquad \forall n\in \N , 
\ee 
thus showing that $(\bx x)^n$ are basic
polynomials for the operator $D_x$, and, when projected, for
$\Delta_x$. So, we have defined an operator $D_x$ which on
functions of $\bx x$ have the same properties as the normal
derivatives $\px$ on functions of $x$. Hence, we can say that
whatsoever is valid for differential equations can be also valid
for the $D_x$ operators provided we consider instead of functions
of $x$ functions depending on $\bx x$. This is the content of the
{\sl \q umbral correspondence}.  In the case of linear differential 
equations, when
the derivations act linearly on functions, the projection
procedure will transform the operator $D_x$ into $\Delta_{x}$ and
we will get  \q difference equations.

Let us analyze  the meaning of the
basic polynomial operators
$(\bx x)^n$ for the three \q derivatives  operators.
Since
\bea
&& (T_x-1) x=x (q_x  T_x -1),\hskip2cm
 (T_x-1)^{-1}x=x(q_xT_x-1)^{-1},
\\
&& (1-T_x^{-1}) x=x (1-\frac{1}{q_x}  T_x^{-1} ),\hskip1.4cm
 (1-T_x^{-1})^{-1}x=x(1-\frac{1}{q_x}  T_x^{-1})^{-1},
\\
&& (T_x-T_x^{-1}) x=x (q_x  T_x -\frac{1}{q_x}T_x^{-1}1),\quad
 (T_x-T_x^{-1})^{-1}x=x(q_xT_x-\frac{1}{q_x} T_x^{-1})^{-1},
\eea
we get
\bea \nonumber
&& \bx^+ x= \qx ^+ x (1+x\px)(q_xT_x-1)^{-1},\\[0.3cm] \label{3.24}
&& \bx^- x= \qx ^- x (1+x\px)(1-\frac{1}{q_x}  
T_x^{-1})^{-1},\\[0.3cm] \nonumber
&& \bx^\s x= \qx ^\s x (1+x\px)(q_xT_x-\frac{1}{q_x}
T_x^{-1})^{-1}. 
\eea 
These
expressions have an operational character. In order to reduce them to a
function we have to project them by acting $\bx x$ on a
constant. So, taking into account that $T_x\;1 = 1$ and
$T^{-1}_x\; 1=1$, in all the three  cases we have 
\be \bx x \;{1}= x .
\ee 
One can demonstrate by induction  that 
\be 
(\bx x)^n\;{1}=
(\bx x) (\bx x) \cdots (\bx x) \;{1} =\frac{n!}{[n]_q !}x^n,
\qquad \forall n\in \N^{+} ,
\ee 
where \be [n]^+
_q=\frac{q^n-1}{q-1}, \qquad [n]^- _q=\frac{1- q^{-n}}{1-q^{-1}},
\ee \be\qquad [n]^\s
_q=\frac{q}{q^2-1}\frac{q^{2n}-1}{q^n}=
\frac{1}{q^{n-1}}\sum_{k=1}^{n}q^{2k-2}
, \ee and \be [n]_q!=  [n]_q [n -1]_q \cdots [1]_q . \ee
Consequently, if we consider an entire function like the
exponential, we have \be e^{\la \bx x}\; 1=\sum_{n=0}^{\infty}
\frac{\la ^n}{n!}(\bx x)^n 1= \sum_{n=0}^{\infty} {\la
^n}\frac{x^n} {[n]_q!} . 
\ee 
Therefore, by the \q umbral correspondence,
the exponential function becomes 
\be \label{3.31} e^{\la x}=
\sum_{n=0}^{\infty} {\la ^n}\frac{x^n} {n!} \longrightarrow
e^{\la \bx x} 1=\sum_{n=0}^{\infty} {\la ^n}\frac{x^n} {[n]_q!} .
\ee 

Let us consider  the gaussian function, that we will be
using later in this work. It takes the form 
\be \label{3.32} 
e^{-\la (\bx
x)^2}\; 1=\sum_{n=0}^{\infty} \frac{(-\la )^n}{n!}(\bx x)^{2 n} 1=
\sum_{n=0}^{\infty} {(-\la )^n}\frac{(2n)!} {n!}\frac{x^{2n}}
{[2n]_q!} . 
\ee

By considering an arbitrary point $x_0$ and an arbitrary real constant 
$a$ we have
\be \label{3.33}
D_x (\bx x+x_0) ^a = a (\bx x +x_0)^{a-1}.
\ee
The proof of eq. (\ref{3.33}) is based on the idea  that the differential equation
\be
(x +x_0)   \px f= a f ,
\ee
whose solution is $f=(x +x_0) ^a$, can be transformed by the \q umbral 
correspondence into the discrete equation
\be
(\bx x +x_0) D_x (\bx x +x_0)^a = a (\bx x +x_0)^a ,
\ee
which has formally the same solution in power series 
(substituting $x$ by $\bx x$ in the discrete case). Effectively,
\be
( x +x_0)^a =\sum_{n=0}^{\infty} \frac{ x_0 ^{a-n}
\prod_{k=0}^{n-1}(a-k)}{n!} x ^n
\ee
while
\be
(\bx x +x_0)^a  =
\sum_{n=0}^{\infty} \frac{ x_0 ^{a-n}\prod_{k=0}^{n-1}(a-k)}{n!}   
(\bx x) ^n  .
\ee
After projection we get
\be
(\bx x +x_0)^a \; 1 =
\sum_{n=0}^{\infty} \frac{ x_0 ^{a-n}\prod_{k=0}^{n-1}(a-k)}{n!}   
(\bx x) ^n  \; 1
=\sum_{n=0}^{\infty} \frac{ x_0 ^{a-n}\prod_{k=0}^{n-1}(a-k)}{[n]_q!}    
x ^n   .
\ee
$\Box$

\sect{\q Umbral functions}
\label{numerical}

In this section we consider some basic discrete functions obtained by means of the
$q$-umbral method and  discuss their range of validity taking
as reference the corresponding solution of the difference
equation and the continuous corresponding function. As
concrete examples we shall consider as \q derivative
operator two of the \q  derivatives of the previous sections: the right
$\Delta^+$ and the symmetric $\Delta^s$. We present here two of the
functions which appear in the symmetry reduction of the heat equation, the
exponential and gaussian functions  (\ref{3.31}), (\ref{3.32}),
which exemplify the kind of results one can obtain from the \q
umbral calculus for entire functions. These functions are
described, respectively,  by the ordinary differential equations:
 \be \label{4.1}
\frac {d f_e(x)}{d x}   = \lambda f_e(x) 
\ee 
and 
\be \label{4.2}
\frac {d f_g(x)}{d x}   = - 2 \lambda x f_g(x). 
\ee
Eq.(\ref{4.1}) has as solution the function $f_e(x) = a
e^{\lambda x}$ and eq.(\ref{4.2}) has as solution the function
$f_g(x) = b e^{- \lambda x^2}$, where $a$ and $b$ are arbitrary
integration constants.

\subsect{\q Exponential functions}

From  the $q$-umbral correspondence the difference equation satisfied
by the $q$--exponential is (see eq.(\ref{4.1})) 
\be \label{exp}
\Delta E_q( x)=\lambda E_q( x) . 
\ee We will discuss
always the domain $x>0$ for all real $\lambda$. The negative values $x<0$ can be
obtained by changing the sign of $\lambda$.

\noindent
\underline{\it  The right exponential}

From (\ref{qderivadas}) the difference equation (\ref{exp}) becomes 
\be\label{dexp}
E_q( q x)= \big[ 1 +(q-1)\la x\big] E_q( x) .
\ee The
solution can be expressed as a product  
\bea
&&E_q( q^n x_0)=\prod_{j=0}^{n-1}\left[1+(q-1)\la q^j x_0\right] E_q(
x_0), \quad n\in\N^+ ,\label{mas}\\ 
&&E_q( q^{-n}
x_0)=\prod_{j=1}^{n}\frac1{\left[1+(q-1)\la q^{-j} x_0\right]} E_q( x_0),
\quad n\in\N^+ .\label{menos} 
\eea 
We have four different behaviours of the
\q exponential function according to the values of $q$ and $\lambda$, namely
$q\gtrless 1$ and $\lambda \lessgtr 0$.

\noindent
\begin{itemize}
\item {\it  $q>1$}

\begin{itemize}
\item Let us at first consider the case $\la>0$. 
The recurrence of the difference equation
implies that $E_q( x)$ is an (monotonous) increasing function in $x$.

By the \q umbral correspondence the solution of the equation
(\ref{4.1}) can be also obtained by \q umbralizing the series
representation of the exponential function. In such a case we
have \be \label{4.3} \widetilde E_q( x)=
\sum_{k=0}^{\infty}\frac{(\la x)^k(q-1)^k}{\prod_{j=1}^{k}(q^j-1)}.
\ee This solution  converges for all $x>0$, so that it
gives the unique solution (\ref{mas}-\ref{menos}) of
(\ref{dexp}) in all the domain.

\item In the case $\la<0$ we see that the recurrence leads to a decreasing
function as long as $1-(q-1)\la x>0$. However, for further values of $x$,
where $1-(q-1)\la x<0$, the function oscillates with higher
diverging amplitudes. For the particular point $x_0=1/((q-1)|\la|)$ we have
$E_q(q x_0)=0$, therefore also $E_q( q^n x_0)=0, \forall n\in \N$. 
This means that in the $q$--lattice $x_n=x_0 q^n, n\in \Z$, the
$q$-exponential decreases for $n$ negative, i.e. for values of $x$ less
than  $x_0$ and vanishes after, for $n$ positive,  avoiding the oscillations.
\end{itemize}

As $q\to 1$ the point $x_0 \to \infty$, so that the $q$--exponential in
that lattice becomes closer and closer to the (continuous)
exponential. The \q umbral function (\ref{4.3}) displays these same
features as it is shown in Fig.\ 1.

\begin{center}
\epsfig{file=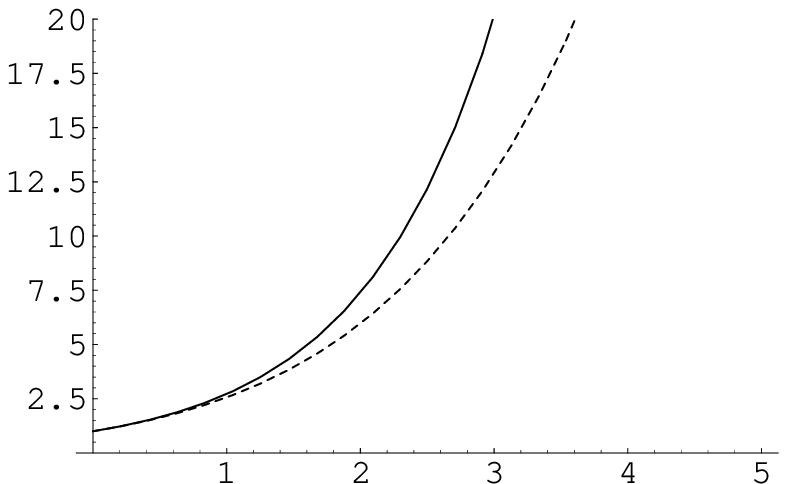, width=10cm}
\vskip0.15cm
\begin{minipage}{13cm}
\small{Fig. 1A. Plotting of the exponential function (continuous line) 
and right $q$-exponential function (dashed
line) for $\la=1$ and $q=1.3$.}
\end{minipage}
\end{center}
\vskip0.25cm

\begin{center}
\epsfig{file=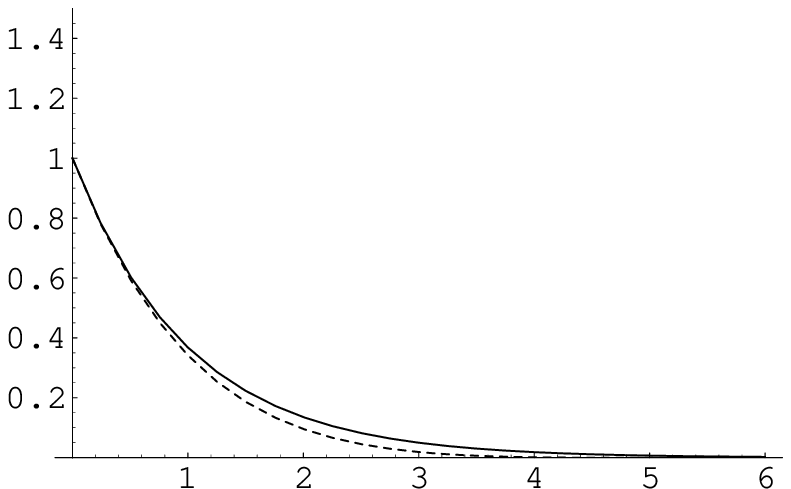, width=10cm}
\vskip0.15cm
\begin{minipage}{13cm}
\small{Fig. 1B. Plotting of the exponential function (continuous line) 
and right $q$-exponential function (dashed
line) for $\la=-1$ and $q=1.3$.}
\end{minipage}
\end{center}
\vskip0.25cm

\begin{center}
\epsfig{file=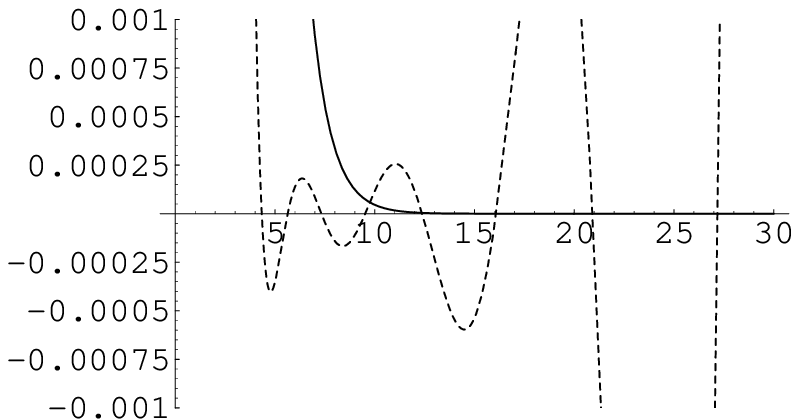, width=10cm}
\vskip0.15cm
\begin{minipage}{13cm}
\small{Fig. 1C. Enlarged plotting of the exponential function 
(continuous line) and right q-exponential function
(dashed line) for $\la=-1$ and $q=1.3$.}
\end{minipage}
\end{center}

\noindent
\item {\it  $q<1$}

\begin{itemize}
\item For $\la>0$ and for big values of $n$ one can always find  
$\big[ 1 +(q-1)\la q^n x_0\big]>0$ and the solution is a monotonous
increasing function for small values of $x_0$. For the value $x_0$ such that
$\big[ 1 +(q-1)\la x_0\big]=0$, the solution is not defined, in fact it 
diverges in that point, and, consequently, also diverges in the lattice $x_0
q^{-n}$, $n\in \N$. For the rest of the points  $x=x_0 q^j$  such that
$\big[ 1 +(q-1)\la q^j x_0\big]<0$, that is, for the values  $x>x_0$ , the
solution oscillates, changing the sign alternatively, with divergences. The
amplitude of the oscillations tends to zero, as
$x\to +\infty$.

\item For $\la<0$ the solution is a monotonous decreasing function that  
in the limit $x\to 0$ goes to zero.

For the \q umbral series the radius of convergence is given by
$R=(|q-1||\la|)^{-1}$. The \q umbral function $\widetilde E_q( x)$
has two vertical asintotes at the symmetric points $x=\pm R$.
Therefore, inside the range of convergence (where there are no
oscillations) the \q umbral function reproduces correctly the
solution $E_q( x)$, but for $x>|R|$ where it does not
converges it supplies no information about the solution. These
features are illustrated in  Fig.\ 2.
\end{itemize}
\end{itemize}

\begin{center}
\epsfig{file=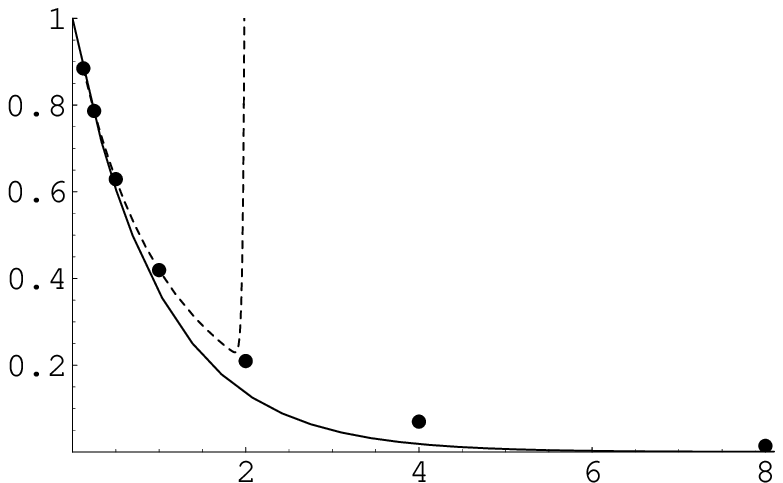, width=10cm}
\vskip0.15cm
\begin{minipage}{13cm}
\small{Fig. 2A. Plotting of the exponential function (continuous line) 
and right $q$-exponential function (dashed
line) and the solution of eq. (\ref{dexp}) given by the points, 
for $\la=-1$ and $q=0.5$.}
\end{minipage}
\end{center}
\vskip0.25cm

\begin{center}
\epsfig{file=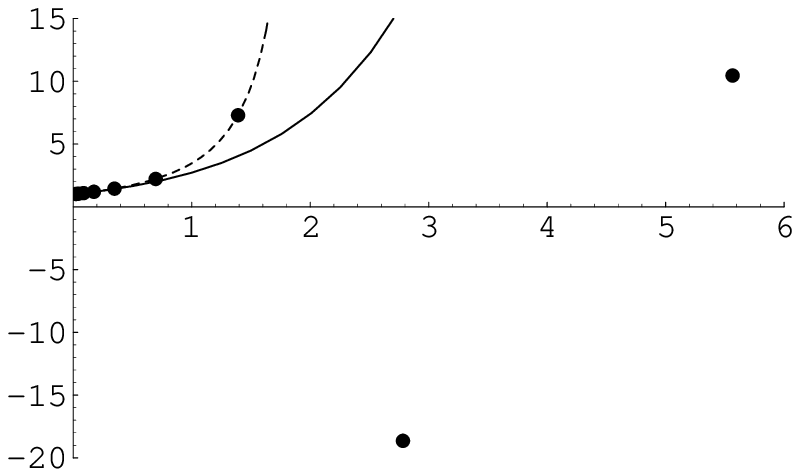, width=10cm}
\vskip0.15cm
\begin{minipage}{13cm}
\small{Fig. 2B. Plotting of the exponential function (continuous line) 
and right $q$-exponential function (dashed
line) and the solution of eq. (\ref{dexp}) given by the points, 
for $\la=1$ and $q=0.5$.}
\end{minipage}
\end{center}
\vskip0.25cm

\begin{center}
\epsfig{file=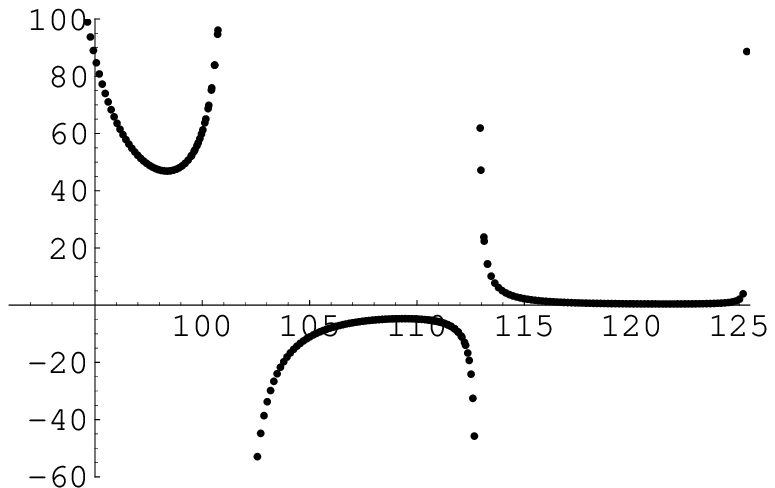, width=10cm}
\vskip0.15cm
\begin{minipage}{13cm}
\small{Fig. 2C. Enlarged plotting of the solution of eq. (\ref{dexp}), 
for $\la=1$ and
$q=0.5$.}
\end{minipage}
\end{center}

\noindent
\underline{\it  The symmetric exponential}

When we consider the $\Delta^{s}$ operator (\ref{3.3}),
the difference equation (\ref{exp}) becomes a three
term relation given by \be \label{sim} E_q^s( q x) = (q-1/q)
\la x E_q^s( x)+ E_q^s(  x/q) .\ee 
So, there are two independent solutions
for the symmetric $q$-exponential. It is  not necessary to
distinguish the cases $q>1$ and $q<1$ because they play a
symmetric role. For $\la>0$ relation (\ref{sim}) gives growing
functions in the $x$ variable. For $\la<0$ the function initially
is decreasing but after a certain point the discrete solutions
start  to oscillate.

The \q umbral solution to the recurrence equation (\ref{sim}) is given by
\be\label{esim} \widetilde E_q^s( q x)=
\sum_{k=0}^{\infty}\frac{(\la
x)^k(q-1/q)^k}{\prod_{j=1}^k(q^j-1/q^j)} . 
\ee 
This series converges
for all $x$, so that it provides one of the solutions to the
recurrence (\ref{sim}), and in the limit $q\to 1 $ it goes into the
continuous exponential. The behaviour of the umbral solution
$\widetilde E_q^s$ presents the features of the recurrence above
described, i. e., it approaches the continuous exponential up to
the first zero (for $\la<0$) but it becomes wildly oscillating
beyond that point (see  Fig.\ 3).

\begin{center}
\epsfig{file=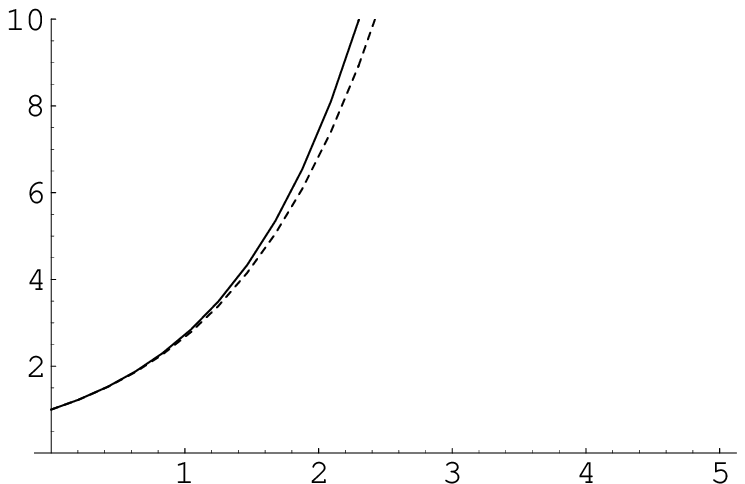, width=10cm}
\vskip0.15cm
\begin{minipage}{13cm}
\small{Fig. 3A. Plotting of the exponential function (continuous line) 
and symmetric $q$-exponential function
(dashed line), for $\la=1$ and $q=1.3$.}
\end{minipage}
\end{center}
\vskip0.25cm

\begin{center}
\epsfig{file=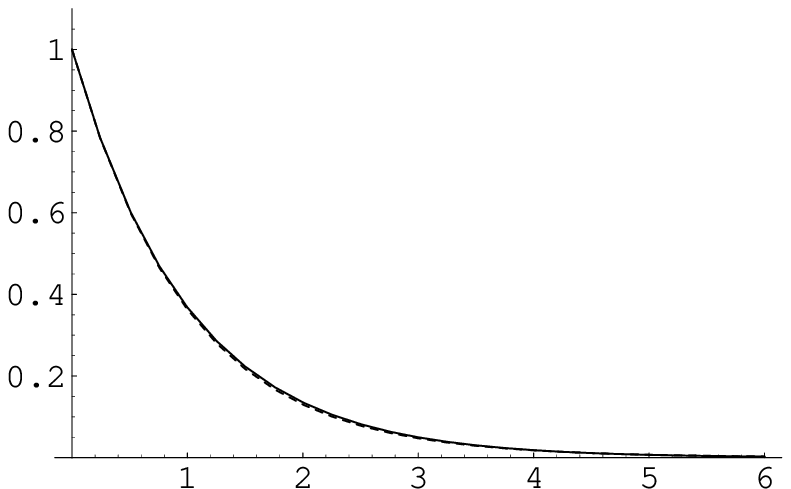, width=10cm}
\vskip0.15cm
\begin{minipage}{13cm}
\small{Fig. 3B. Plotting of the exponential function (continuous line) 
and symmetric $q$-exponential function
(dashed line), for $\la=-1$ and $q=1.3$.}
\end{minipage}
\end{center}
\vskip0.25cm

\begin{center}
\epsfig{file=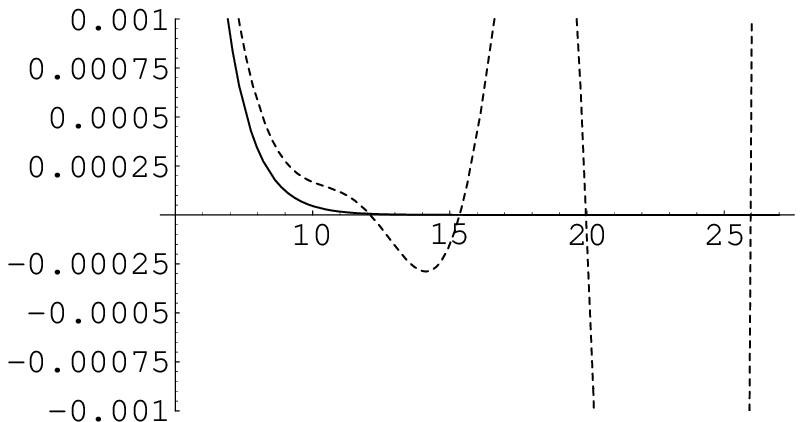, width=10cm}
\vskip0.15cm
\begin{minipage}{13cm}
\small{Fig. 3C. Enlarged plotting of the exponential function 
(continuous line) and symmetric $q$-exponential
function (dashed line), for $\la=-1$ and $q=1.3$.}
\end{minipage}
\end{center}

Comparing Fig.\ 1 and  Fig.\ 3 we can see that the symmetric
exponential function gives a better approximation than the right 
exponential. However, it can be easily shown that there is no initial
condition for equation (\ref{sim}{\bf )} such that the symmetric
$q$--exponential vanishes for all subsequent points, as it was
the case for the right exponential. So, there is no way to avoid
the oscillations.

\subsect{q-Gaussians}

In this case eq.(\ref{4.2}), which has as a solution the gaussian function,  
becomes the \q difference equation
\be\label{ga}
\Delta G_{q,\la}(x) = -\la x \beta_{x} G_{q,\la}(x).
\ee

In the following we will discuss briefly the cases of the right and 
symmetric $q$--gaussians.

\noindent
\underline{\it  The right gaussian}

The difference equation (\ref{ga}) becomes a three--term
recurrence equation 
\be\label{recga} q^{-1}
G_{q,\la}^+(q^2x) - (q^{-1}+1)G_{q,\la}^+(q x) + G_{q,\la}^+(x) =
-\la (q-1)^2x(x^2\partial_x+1) G_{q,\la}^+(x) .
\ee 
However, now eq. (\ref{recga}) is a differential--difference
equation. So, it is quite difficult to find directly the solutions or even
to discuss  the general behaviour of its solutions. We will assume, as shown 
in the case of the exponential function, that, whenever the \q umbral series 
is convergent it will converge to the solution of the difference equation.

The \q umbral series supplies a
solution to eq. (\ref{recga}), given by
\be\label{umga} \widetilde G_{q,\la}^+( x)=
\sum_{k=0}^{\infty}\frac{(-\la x^2)^k 2^k (2k -1)!(q-1)^{2k}}{
\prod_{j=1}^{2k}(q^j-1)} .
\ee 
For $q>1$ the series converges for all
$x$, but  for $q<1$ the series diverges everywhere (for $x\neq
0$). Some plottings of
$\widetilde G_{q,\la}^+$ are shown in  Fig.\ 4.  The behaviour is
similar to that of the $q$--exponential. Whenever  we have a decreasing
function of $x$,  at a certain point it vanishes and beyond that point it
starts to oscillate with increasing amplitudes, thus departing
from the behaviour of the continuous gaussian function.

\begin{center}
\epsfig{file=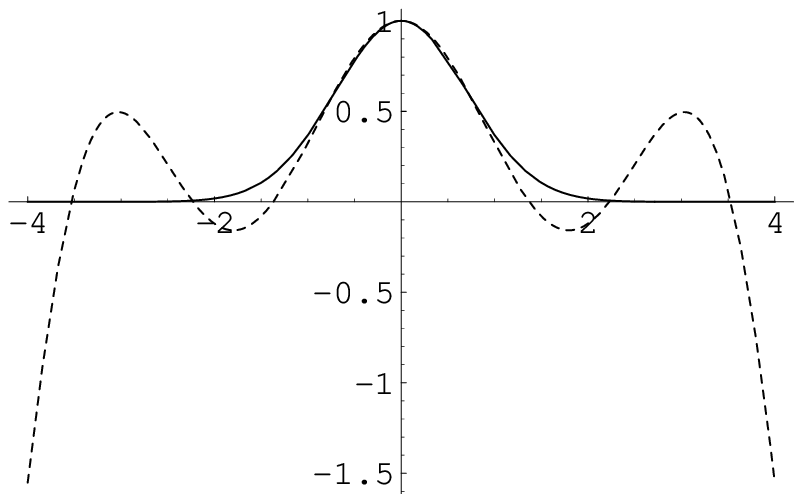, width=10cm}
\vskip0.15cm
\begin{minipage}{13cm}
\small{Fig. 4A. Plotting of the gaussian function (continuous line) 
and right $q$-gaussian function
(dashed line), for $\la=1$ and $q=1.3$.}
\end{minipage}
\end{center}

\vskip0.25cm

\begin{center}
\epsfig{file=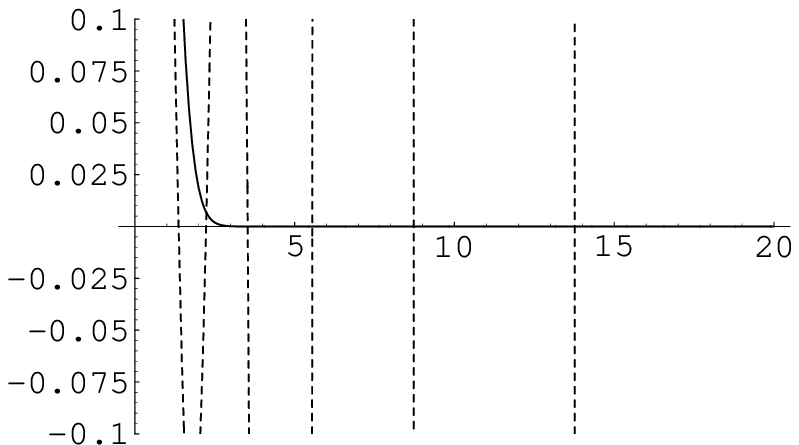, width=10cm}
\vskip0.15cm
\begin{minipage}{13cm}
\small{Fig. 4B. Enlarged plotting of the gaussian function 
(continuous line) and right $q$-gaussian function
(dashed line), for $\la=1$ and $q=1.3$.}
\end{minipage}
\end{center}

\noindent
\underline{\it The symmetric gaussian}

The recurrence relation of eq. (\ref{ga}) for this case is
an even more involved differential difference equation than
eq.(\ref{recga}), so we prefer not to write it down. The 
\q umbral series solution is given by 
\be\label{umgasim} \widetilde
G_{q,\la}^s( x) = \sum_{k=0}^{\infty}\frac{(-\la
x^2)^k(2k)!(q-1/q)^{2k}}{k! \prod_{j=1}^{2k}(q^j -1/q^j)} .
\ee 
The radius of convergence is $R=\infty$. For the symmetric gaussian
function we have similar results as for the right gaussian, which, however,
are valid for any value of $q$ (see Fig.5).

\begin{center}
\epsfig{file=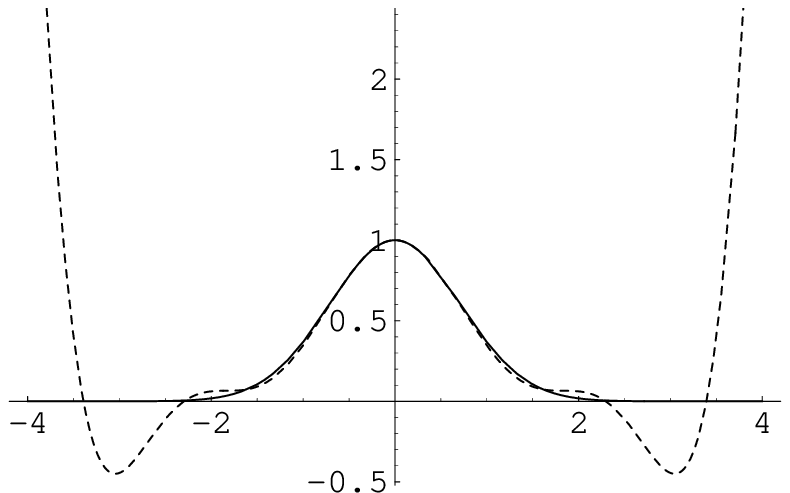, width=10cm}
\vskip0.15cm
\begin{minipage}{13cm}
\small{Fig. 5A. Plotting of the gaussian function (continuous line) 
and symmetric $q$-gaussian function
(dashed line), for $\la=1$ and $q=1.3$.}
\end{minipage}
\end{center}

\vskip0.25cm

\begin{center}
\epsfig{file=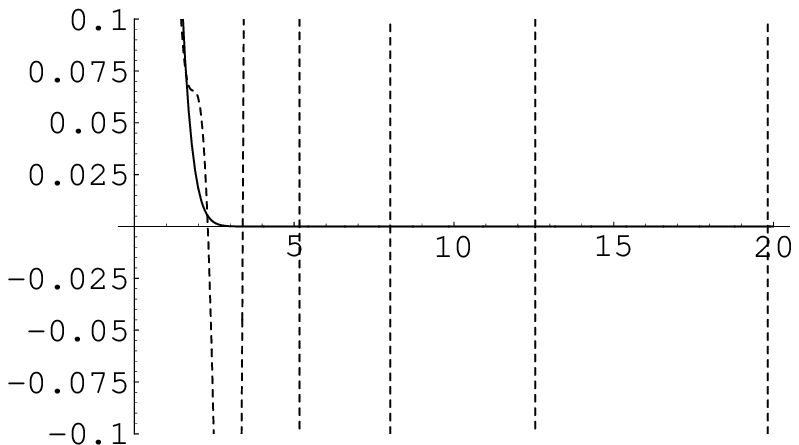, width=10cm}
\vskip0.15cm
\begin{minipage}{13cm}
\small{Fig. 5B. Enlarged plotting of the gaussian function 
(continuous line) 
and symmetric $q$-gaussian function
(dashed line), for $\la=1$ and $q=1.3$.}
\end{minipage}
\end{center}

In conclusion, we can say
that the decreasing asymptotic behaviour of the classical
functions is not fully reproduced by the corresponding umbral
$q$--functions. In that region the discrete functions approach
the classical ones up to a point where they vanish, but beyond
this point they oscillate going far away from the continuous
 analogues. Therefore, a good parameter
measuring the radius of the
domain where the $q$--functions imitate the continuous functions
is given by the first zero in the region of asymptotic behaviour.
This is depicted in Fig. 6 for the $q$--exponentials and the
$q$--gaussians.

From these plottings we evince that the domain of convergence of 
the \q exponential function to the exponential function increases 
in a monotonic continuous way as $q \rightarrow 1$ for the right 
exponential while in the symmetric case there are discontinuities 
for small values of $q$. 
The situation is slightly different in the case of the gaussian 
function, when, as $q$ decreases also the domain decreases up to 
a minimum $\la$ dependent value $q_0$. Below $q_0$
the domain increases as $q \rightarrow 1$. In the case of the symmetric
gaussian function we again have that for $q$ small the function is discontinuous.

\begin{center}
\epsfig{file=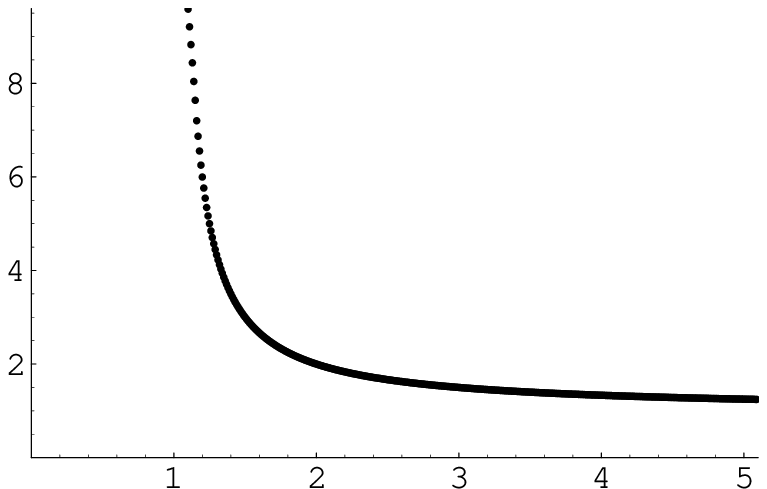, width=10cm}
\vskip0.15cm
\begin{minipage}{13cm}
\small{Fig. 6A. Plotting of the position of first zero of the right 
$q$-exponential as function of $q$ for $\la=-1$}
\end{minipage}
\end{center}

\begin{center}
\epsfig{file=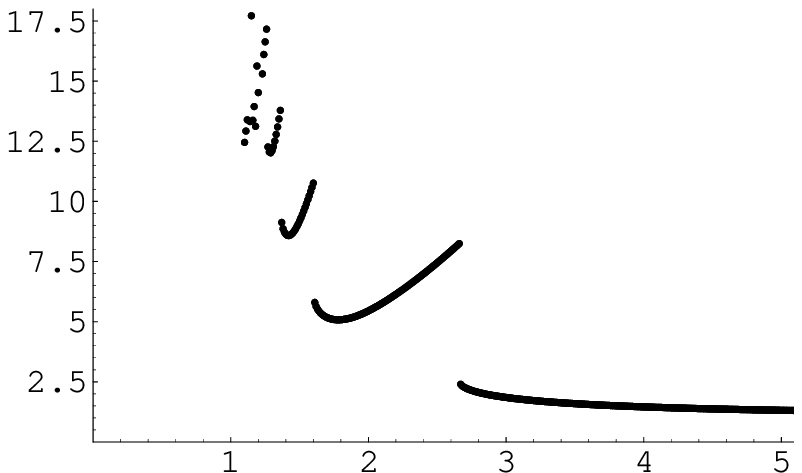, width=10cm}
\vskip0.15cm
\begin{minipage}{13cm}
\small{Fig. 6B. Plotting of the  position of first zero of the symmetric 
$q$-exponential as function of $q$ for $\la=-1$}
\end{minipage}
\end{center}

\begin{center}
\epsfig{file=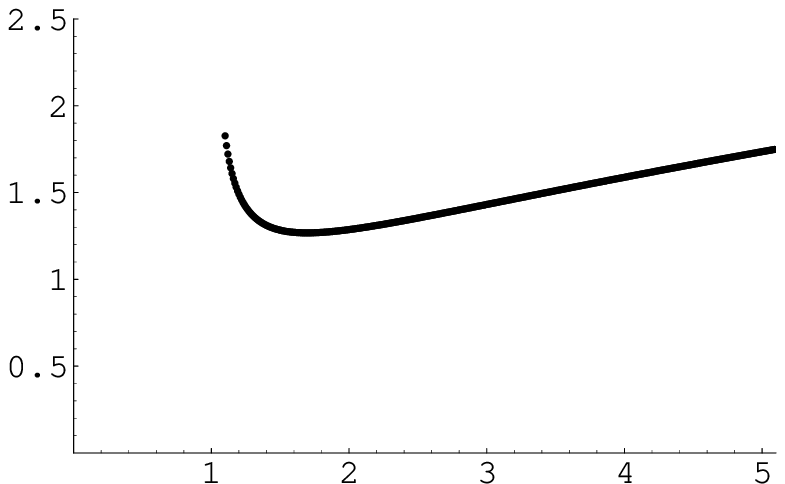, width=10cm}
\vskip0.15cm
\begin{minipage}{13cm}
\small{Fig. 6C.  Plotting of the  position of first zero of the right 
$q$-gaussian as function of $q$ for $\la=1$}
\end{minipage}
\end{center}

\begin{center}
\epsfig{file=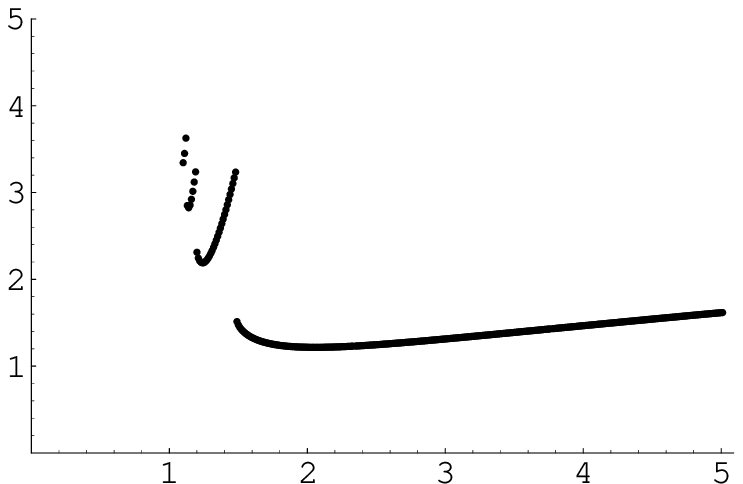, width=10cm}
\vskip0.15cm
\begin{minipage}{13cm}
\small{Fig. 6D.  Plotting of the  position of first zero of the symmetric 
$q$-gaussian as function of $q$ for $\la=1$}
\end{minipage}
\end{center}

\sect{Examples}
\label{examples}

\subsect{Discrete heat \q equation}
\label{heatequation}
Taking into account the $q$-umbral correspondence
\be \label{5.0}
\px\to \Delta_{x} , \qquad x\to \bx x
\ee
we obtain from any linear differential equation with constant coefficients an 
operator equation which,
when projected, gives us a \q discrete
  equation. In the case of the heat equation we get
\be \label{5.1}
(\pt -\pxx ) u=0 \quad \Longrightarrow \quad  
(\Delta_{t}-\Delta_{xx}) u=0
\ee
Let us consider now the problem of obtaining the symmetries for the
\q discrete eq.(\ref{5.1}). We can apply the \q
umbral correspondence also to the determining equations, as they are 
linear in the infinitesimal coefficients
$\xi$, $\tau$ and $f$. So, making use of the Leibniz rule we obtain the  
following set of determining equations:
\bea\label{ecuacacionesdeterminantescalor}
&& D_{x}(\tau)=0 ,\\[0.2cm]
&& D_{t}(\tau) -2 Q^{x}(\xi)  =0 ,\\[0.2cm]
&& D_{t}(\xi)  - Q^{xx}(\xi)  -2 Q^{x}(f) =0 ,
\\[0.2cm]
&& D_{t} (f)  -Q^{xx}(f)=0 , \eea 
where 
$D_{x}(\tau) 1 = [\Delta_x , \tau ] 1 = \Delta_x \tau$. Moreover 
$$
D_{xx}(f) 1 =D_{x}(D_{x}(f)) 1
=[\Delta_{x},[\Delta_{x},f]] 1 = \Delta_{x} \Delta_{x} f .
$$

Note that these determining equations have formally the same
expression for all \q derivatives,  for the continuous
derivatives and  in the discrete case studied in \cite{lno01}.
From eq.(\ref{5.0}), the solution of  this system
(\ref{ecuacacionesdeterminantescalor})  is
\bea\label{calorsolucionesdeterminingequation} && \tau=\tau_2 (\bt t)^{2} +
\tau_1 (\bt t)+\tau_0 ,\\[0.2cm] && \xi=\frac 12 (\tau_1+2\tau_2 (\bt
t))(\bx x)+\xi_1 (\bt t)+\xi_0, \\[0.2cm] && f=\frac 14  \tau_2 (\bx x
)^2+\frac 12  \tau_2 (\bt t)+
\frac 12 \xi_1 (\bx x)+\gamma ,
\eea
where $\tau_0,\  \tau_1,\  \tau_2,\  \xi_0,\ \xi_1$ and $\gamma$ are
arbitrary functions of $T_x$, $T_t$ and of $q_x$ and $q_t$.

By a suitable choice of the functions  
$\tau_1,\ \tau_2,\ \tau_0,\ \xi_1, \ \xi_0$ and
$\gamma$ we get  the following representation of the symmetries:
\bea\label{caloralgebra}
&& P_0^{q} = (\Delta_t   u)\partial_u  ,\\[0.2cm]
&& P_1^{q} =(\Delta_x   u)\partial_u , \\[0.2cm]
&& W^{q} = u\partial_u , \\[0.2cm]
&& B^{q} =(2(\bt t) \Delta_t  u  +(\bx x )\Delta_x  u )\partial_u  ,
\\[0.2cm]
&& D^{q} = (2 (\bt t) \Delta_t  u +(\bx x)\Delta_x  u + 
\frac 12 u)\partial_u ,
\\[0.2cm]
&& K^{q} =((\bt t)^2\Delta_t  u  + (\bt t)(\bx x)\Delta_x  u  +  
\frac 14 (\bx x)^2  u  + \frac12 (\bt t )u)\partial_u ,
\eea
that close into a 6--dimensional Lie algebra, isomorphic to the
symmetry algebra of the continuous heat equation.
Another realization of this algebra was obtained in    
\cite{javier}, by a different
procedure and  used to find symmetric solutions of the
discrete heat equation.

Now we use these symmetries to construct few solutions of the 
\q discrete equation
(\ref{5.1}).
Taking into account the symmetries $P_0^{q}$ and $P_1^{q}$,
 choosing the \q parameters in such a way that $\qx =q_t = q$ and, 
using the variable separation method
\cite{miller}, we can write the solution of the \q
heat equations as 
\be
 u(\bt t,\bx x)=v(\bt t) w(\bx x) .
\ee 
Then
eq.(\ref{5.1}) reads 
\be \label{5.1.1} (\Delta_{t} v(\bt t) )
w(\bx x) - v(\bt t) (\Delta_{xx}) w(\bx x)) = 0. \ee 
From
eq.(\ref{5.1.1}) we deduce that the functions $v$ and $w$ must
satisfy the following equations 
\be \label{5.1.2} \Delta_{t}
v(\bt t) = \la v(\bt t), \qquad \Delta_{x} w(\bx x) = \sqrt{\la}
w(\bx x) . 
\ee 
By means of eq.(\ref{exp}), we have  \be v(t) =
v(\bt t) 1 = e^{\la \bt t} 1=
 \sum_{n=0}^{\infty} \frac{\la^n  t^{n}} {[n]_q!}= e_q^{\la t }
\ee 
and 
\be w(x) = w(\bx x) 1 =
e^{\sqrt{\la} \bx x} 1=
 \sum_{n=0}^{\infty} \frac{\la^{n/2}  x^{n}} {[n]_q!}= 
e_q^{\sqrt{\la} x }.
\ee Hence, the solution will be 
\be u(t,x) = e_q^{\la  t}
e_q^{\sqrt{\la} x}. 
\ee 

Let us consider now the symmetry
reduction with respect to the operator $B^{q}$ \cite{olver}. In this case
introducing the appropriate symmetry variable $\eta = \frac{\bx
x}{\sqrt{\bt t}}$ we get 
\be \label{5.1.5}
 u(x,t) =
\frac{u_0}{\sqrt{\bt t}} \exp [ - \frac{(\bx x)^2}{ 4 \bt t} ]\, 1 .
\ee 
The solution (\ref{5.1.5}) of the \q heat equation is
meaningful as long as we are considering positives times and the
value $t=0$ is out of our time domain. In such a situation the
solution (\ref{5.1.5}) is entire and can be represented as a
Taylor series and, thus, \q functions like the gaussian or the
square root are meaningful. The method would  not provide a
meaningful \q function if we would consider all values of $t$.
However, in $t=0$ also the  boost solution of the continuous heat
equation  would be singular and, thus, meaningless.

\subsect{A generalized Hermite equation}
\label{secondexample}

Let us consider the following \q difference equation
\be \label{5.2.1}
\Delta_{xx} \psi + \bx x \Delta_x \psi = E \psi.
\ee
 By explicitating the operator $\bx$ appearing in eq.(\ref{5.2.1}) 
according to (\ref{tres.doce.bis}),
we can rewrite it as
\be \label{5.2.2}
\Delta_{xx} \psi + x \psi_x  = E \psi.
\ee
By the \q umbral correspondence eq.(\ref{5.2.1}) is related to the 
ordinary differential equation $\psi_{xx}
+ x \psi_x = E \psi$ whose solutions are given in terms of gaussian 
functions and Kummer confluent
hypergeometric functions. Explicitly we have \cite{as}:
\bea \label{5.2.3}
\psi(x) &=& A_1 e^{- \frac{x^2}{2}} M( \frac{1}{2} + \frac{1}{2} E, 
\frac{1}{2}, \frac{1}{2} x^2) +
 A_2 x e^{- \frac{x^2}{2}} M( 1 + \frac{1}{2} E, \frac{3}{2}, 
\frac{1}{2} x^2) \\ \nonumber
&=&  e^{- \frac{x^2}{2}} \phi(x) .
\eea
 Both the gaussian and the
Kummer confluent hypergeometric functions have a power series
expansions in terms of $x$, 
\bea 
\phi(x) = A_1 [ 1 + (1 +
\frac{1}{2} E) \frac{x^2}{2!} + ( 1 + \frac{1}{2} E) (3 +
\frac{1}{2} E) \frac{x^4}{4!}  + \dots ] + \\ \nonumber A_2 [ x +
(2 + \frac{1}{2} E) \frac{x^3}{3!} + ( 2 + \frac{1}{2} E) (4 +
\frac{1}{2} E) \frac{x^5}{5!}  + \dots ] .
\eea 
So, the \q differential difference equation (\ref{5.2.2}) will have a
solution in terms of \q gaussian and \q Kummer confluent
hypergeometric functions. Let us notice that, if $E = -2 n$ with
$n$ a positive   integer number, the \q Kummer confluent
hypergeometric function is just a polynomial.

\sect{Conclusions}\label{conclusiones}

In this paper we present a \q extension of the umbral calculus and use 
it to provide solutions of linear \q
difference and \q differential difference equations. In this way we 
obtain solutions which have
a continuous limit.

The discretization procedure given by the recipe
$\px \longrightarrow \Delta_{x}$ and $ x \longrightarrow \bx {x}$
  works well for linear equations also in the case of \q shifts operators. 
In particular, it preserves the
classical Lie symmetries which are described by linear equations.

We study in detail the behaviour of the \q exponential and 
\q gaussian functions and show their range of
validity which depends from the \q discrete delta operator under 
consideration. The domain of convergence of the \q function to the 
continuous function is characterized in term of the zeroes 
 of the \q function.
The results are usually
better in the case of symmetric \q delta operators.

Further work is in process on the complete description of a coherent 
\q umbral calculus and a comparison of the
discrete and $q$--discrete solutions.

\section*{Acknowledgments}
This work has been partially supported by
the DGI of the Ministerio de Ciencia y 
Tecnolog\'{\i}a of Spain (project BMF2002-02000) 
and the Programme FEDER of the European
Community,  and the Junta de Castilla y Le\'on (Spain).  The
visit of DL to Valladolid and of MO to Roma have been partially financed by
the Erasmus European Project.


\end{document}